\documentclass[a4paper]{jpconf}

\usepackage{graphicx,times}
\usepackage{subfigure}

\usepackage{booktabs}

\usepackage{enumerate}
\usepackage{amsmath}
\usepackage{cases}

\usepackage{hyperref}
\usepackage{epstopdf}
\usepackage{amsmath,bm}
\usepackage{amssymb}

\usepackage{morefloats}
\usepackage{multirow}
\usepackage{array}
\usepackage{verbatim}
\usepackage{setspace}

\begin{document}
\title{Synchrotron spectra of GRB prompt emission and pulsar wind nebulae}

\author{Siyao Xu}

\address{Department of Astronomy, University of Wisconsin, 475 North Charter Street, Madison, WI 53706, USA; Hubble Fellow}
\ead{sxu93@wisc.edu}


\begin{abstract}
Particle acceleration is a fundamental process in many high-energy astrophysical environments 
and determines the spectral features of their synchrotron emission.
We have studied the adiabatic stochastic acceleration (ASA) of electrons arising from the basic dynamics of 
magnetohydrodynamic (MHD) turbulence and found that the ASA acts to efficiently harden 
the injected electron energy spectrum. 
The dominance of the ASA at low energies and the dominance of synchrotron cooling at high energies
result in a broken power-law shape of both electron energy spectrum and photon synchrotron spectrum. 
Furthermore, we have applied the ASA to studying the synchrotron spectra of the prompt emission of gamma-ray bursts (GRBs)
and pulsar wind nebulae (PWNe). 
The good agreement between our theories and observations confirms that the stochastic particle acceleration is indispensable in explaining 
their synchrotron emission. 
\end{abstract}

\section{Introduction}

Many cosmic accelerators are huge reservoirs of magnetic energy. 
The dissipated magnetic energy is converted to energies of particles, accounting for the observed synchrotron emission
\cite{Zh11,Deng15,Laz18}. 
In both turbulent and magnetized medium, a proper description of magnetohydrodynamic (MHD) turbulence is crucial for studying 
the interaction between particles and 
turbulent magnetic fields and the resulting particle acceleration 
\cite{XY13,Xuc16,Xult}.
Recent theoretical 
\cite{GS95, LV99} 
and numerical 
\cite{CV00,MG01,CLV_incomp}
studies reveal a critical balance between the turbulent motions in the direction perpendicular to the local magnetic field 
and magnetic wave-like motions parallel to the local magnetic field in MHD turbulence. 
Accordingly, the turbulent dynamo with magnetic field lines stretched by turbulent motions 
\cite{XL16}
and the turbulent reconnection of stochastic magnetic fields
\cite{LV99}
are also in dynamical balance in MHD turbulence. 
As a new acceleration mechanism proposed by 
\cite{BruL16}, 
particles entrained on turbulent magnetic field lines undergo cycles of deceleration in turbulent dynamo regions
and acceleration in turbulent reconnection regions, leading to a globally diffusive acceleration process, 
which we term as 
``adiabatic stochastic acceleration (ASA)''. 
The ASA becomes the dominant acceleration process in MHD turbulence when the non-adiabatic
resonant scattering of particles by anisotropic MHD turbulence is inefficient 
\cite{YL02}. 
In our recent studies, 
we applied the ASA to interpreting 
the Band function spectrum 
\cite{Band93}
of the prompt emission of gamma-ray bursts (GRBs) 
\cite{XZg17,XYZ18}
and the synchrotron spectra of pulsar wind nebulae (PWNe)
\cite{Xu19}.

\section{Electron energy spectrum resulting from the ASA}

The time evolution of the electron energy spectrum $N(E,t)$ is described by 
\begin{equation}\label{eq: soasi}
   \frac{\partial N}{\partial t} =  a_2 \frac{\partial}{\partial E} \Big(E\frac{\partial (EN)}{\partial E}\Big)  
    + \beta \frac{\partial (E^2N)}{\partial E} + Q(E).
\end{equation}
The three terms on the RHS correspond to the ASA, radiation losses, and particle injection. 
The acceleration rate of ASA is 
\begin{equation}\label{eq: a2}
    a_2 = \xi \frac{u_\text{tur}}{l_\text{tur}},
\end{equation}
where $l_\text{tur}$ and $u_\text{tur}$ are the correlation length and speed of turbulence, 
$\xi = \Delta E / E$ is the cumulative fractional energy change during the trapping of particles within individual turbulent eddies during the eddy turnover time 
$\tau_\text{tur} = l_\text{tur}/u_\text{tur}$.
$\xi$ is of order unity for non-relativistic turbulence and of order $\gamma_\text{tur}^2$ for relativistic turbulence with 
$\gamma_\text{tur}$ as the turbulence Lorentz factor.
In the case of both 
synchrotron and synchrotron-self-Compton (SSC) losses, 
$\beta$ is expressed as 
\begin{equation}\label{eq: betp}
    \beta 
               = \frac{\sigma_T c B^2 (1+Y)}{ 6 \pi (m_e c^2)^2},
\end{equation}
where $Y$ is the ratio between the powers of SSC and synchrotron radiation and is zero when only synchrotron is considered. 
Besides, $B$ is the magnetic field strength, 
$\sigma_T$ is the Thomson cross section, 
$\gamma_e$ and $m_e$ are the electron Lorentz factor and the electron rest mass, and 
$c$ is the speed of light. 
The third term $Q(E) = C E^{-p}$ represents a steady injection of power-law electron spectrum with a power-law index $p$, 
accounting for other possible instantaneous acceleration processes, e.g., shock acceleration, reconnection acceleration, 
that generate power-law electron spectra.

\begin{figure}[h]
\includegraphics[width=9cm]{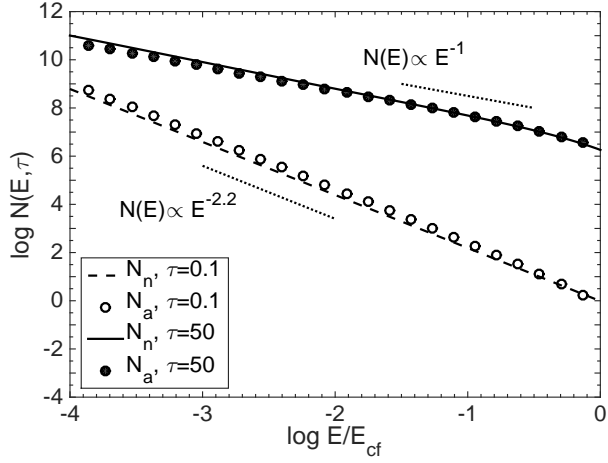}\hspace{1pc}%
\begin{minipage}[b]{14pc}
\caption{ The electron energy spectrum at different times. From \cite{XYZ18}.}
\label{fig: num}
\end{minipage}
\end{figure}

In \cite{XZg17,XYZ18}, we analytically solved Eq. \eqref{eq: soasi} in the energy range where the ASA dominates over radiation losses. 
We found that $N(E,t)$ evolves from a spectral shape governed by the injected particle distribution 
\begin{equation}\label{eq: nsht}
   N(E,\tau)  =  \frac{C^\prime \sqrt{\tau}}{\sqrt{\pi}} E^{-p} \exp \Big(-\frac{E}{E_\text{cf}}\Big),
\end{equation}
to a hard spectrum under the effect of ASA, 
\begin{equation}\label{eq: fism}
     N(E,\tau)   
     =  \frac{C^\prime (E_u^{-p+1} - E_l^{-p+1}) \sqrt{\tau} }{(-p+1)\sqrt{\pi }}    
         E^{-1} \exp \Big(-\frac{E}{E_\text{cf}}\Big)  ,
\end{equation}
where $C^\prime = C/a_2$, $\tau = a_2 t $,  and $E_l $ and $E_u$ are the lower and upper limits of the energy range of injected electrons.  
Here $E_\text{cf} = a_2 / \beta$ is the cutoff energy of the ASA, 
where the acceleration rate equalizes with the cooling rate. 
This hardening of electron energy spectrum due to the ASA is also illustrated in Fig. \ref{fig: num}. 
It shows that irrespective of the injected spectral form, the ASA leads to diffusive particle acceleration and thus 
a hard electron energy spectrum with the power-law index approaching $-1$.

\section{Synchrotron spectrum resulting from both ASA and synchrotron cooling} 
\label{sec: ssasa}

\begin{figure}[h]
\centering
\subfigure[Case (\romannumeral1)]{
   \includegraphics[width=7.5cm]{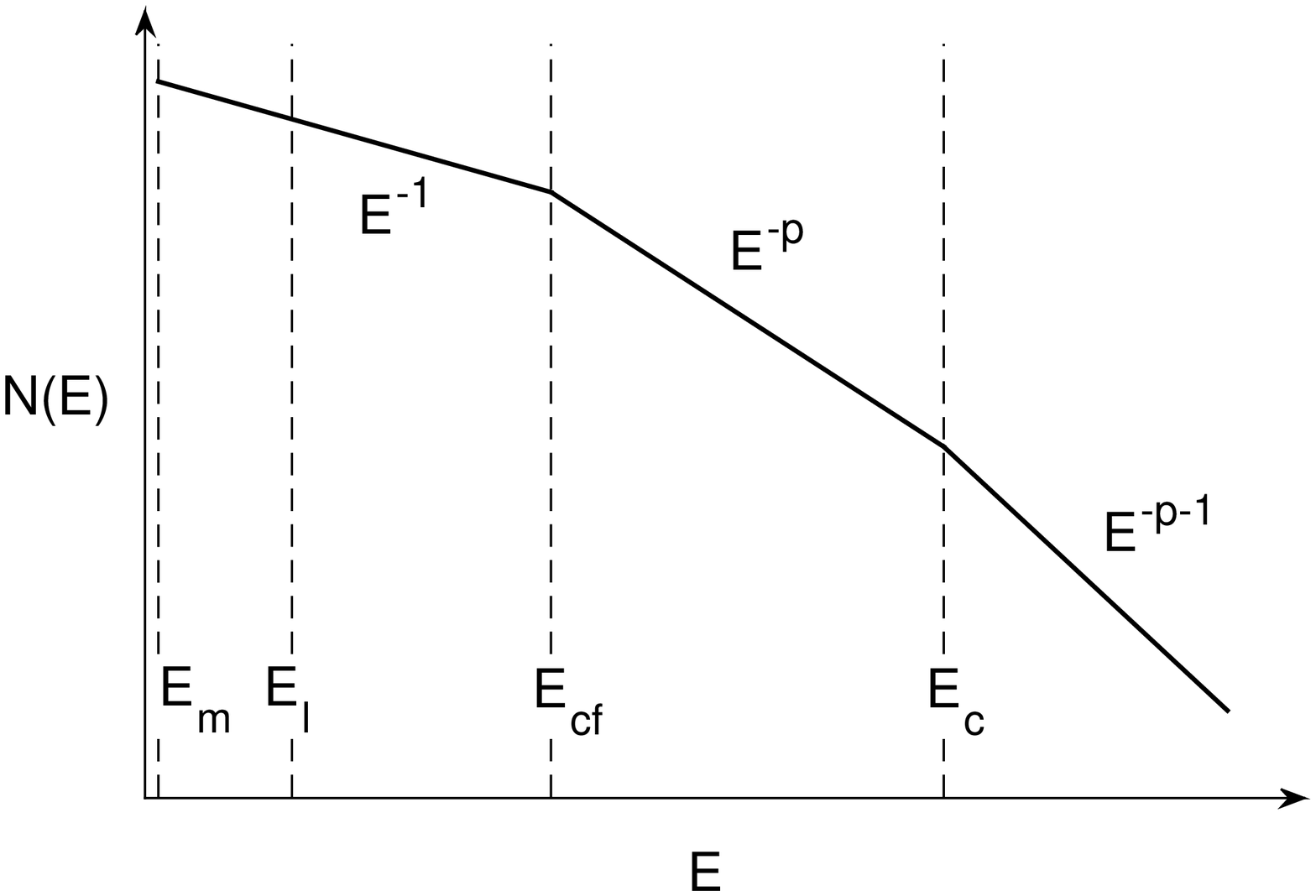}\label{fig: sloc1}}
\subfigure[Case (\romannumeral1)]{
   \includegraphics[width=7.5cm]{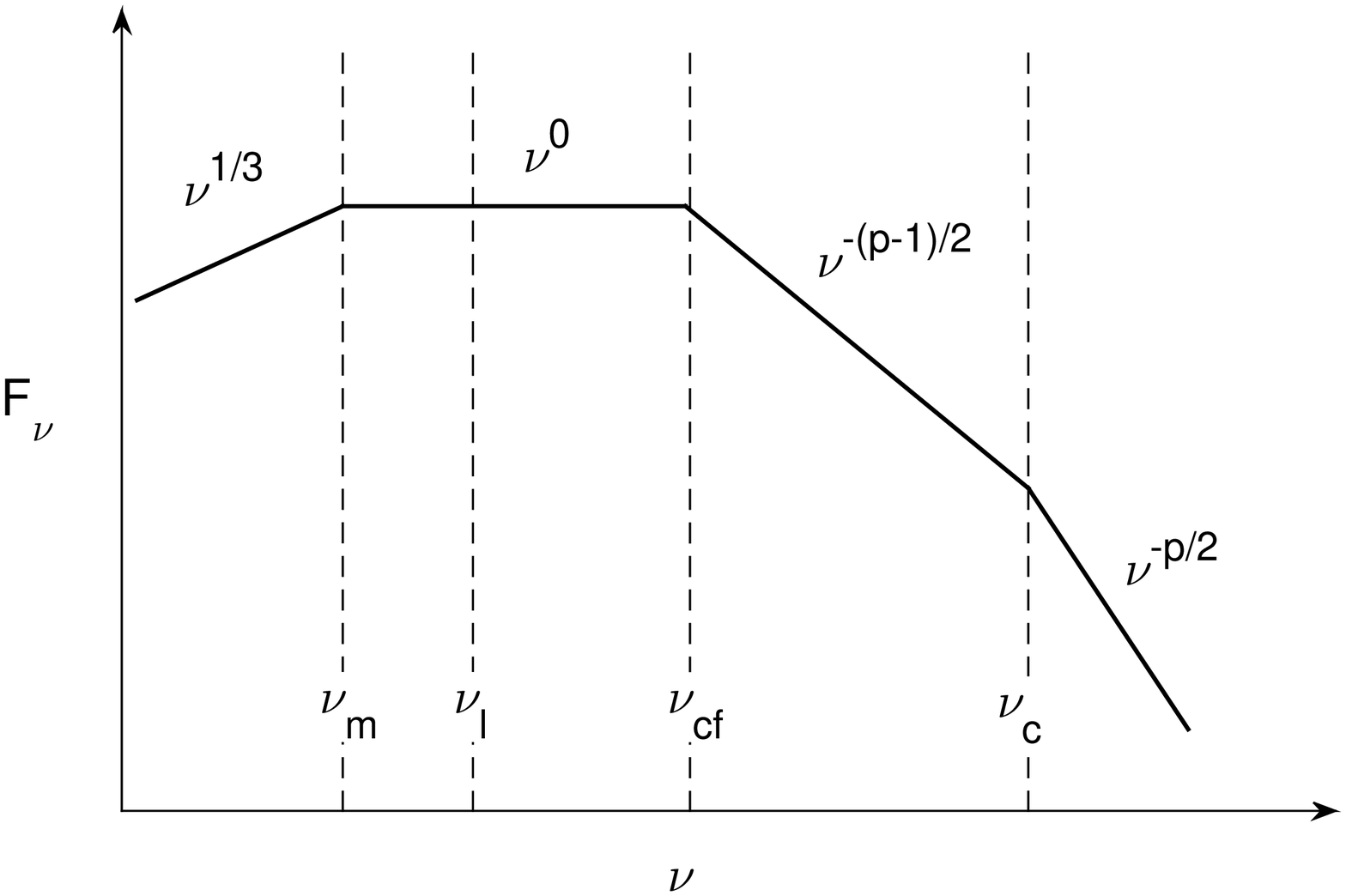}\label{fig: sloc1s}}   
\subfigure[Case (\romannumeral2)]{
   \includegraphics[width=7.5cm]{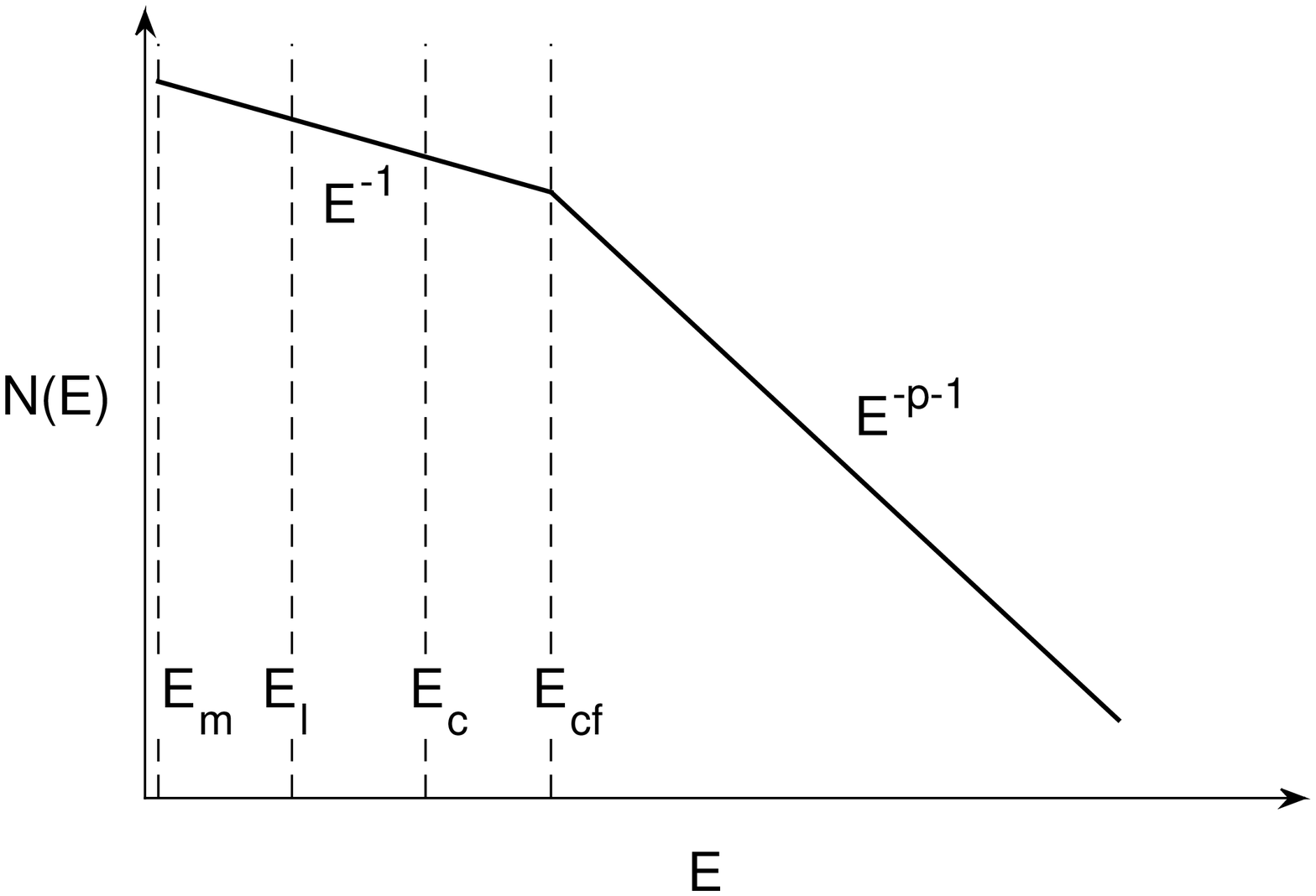}\label{fig: sloc2}}
\subfigure[Case (\romannumeral2)]{
   \includegraphics[width=7.5cm]{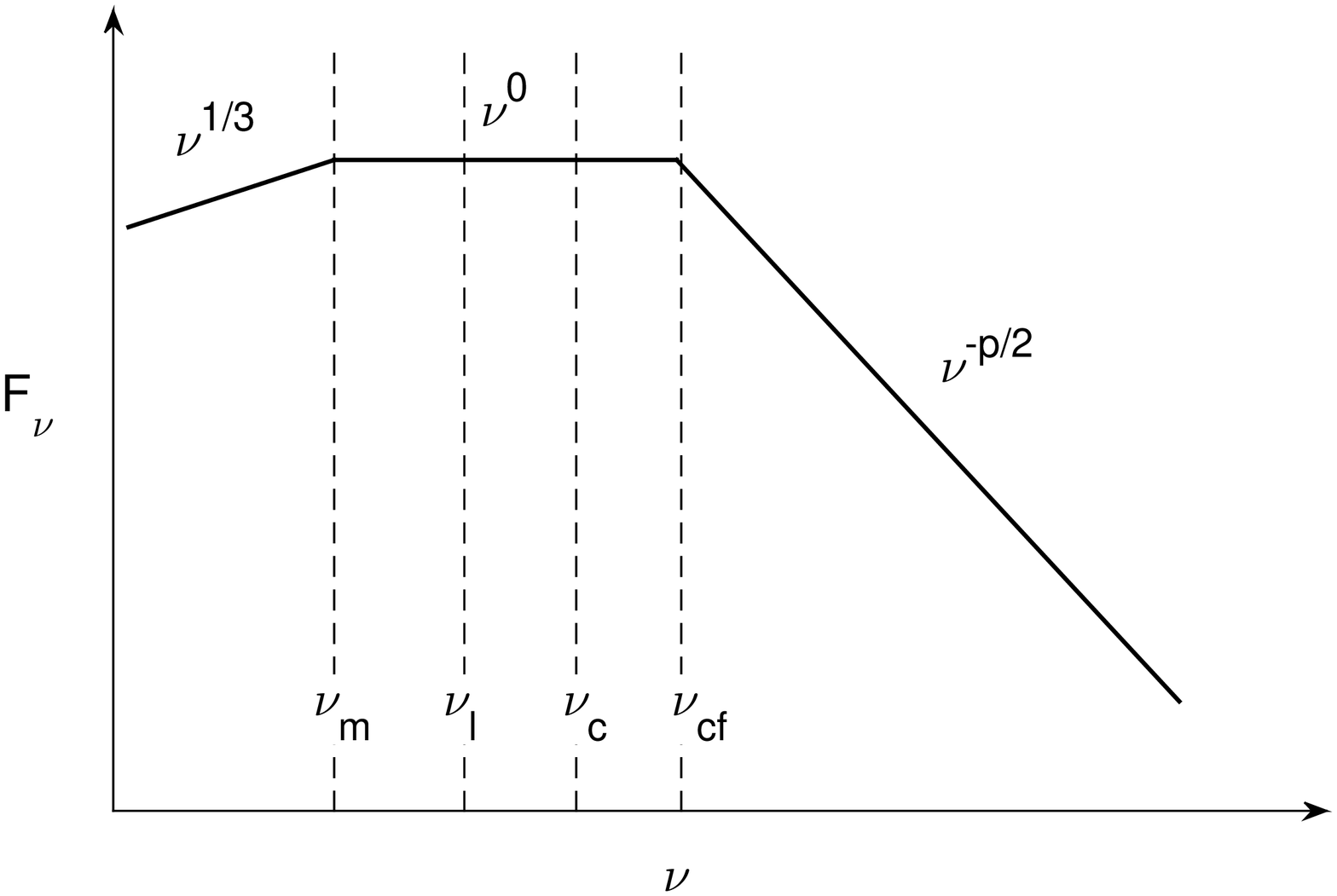}\label{fig: sloc2s}} 
\caption{The electron energy spectrum ((a) and (c)) and the corresponding synchrotron spectrum ((b) and (d)) in Case (\romannumeral1)
and in Case (\romannumeral2). From \cite{XYZ18}. }
\label{fig: fast}
\end{figure}

\begin{figure}[h]
\centering
\subfigure[Case (\romannumeral3)]{
   \includegraphics[width=7.5cm]{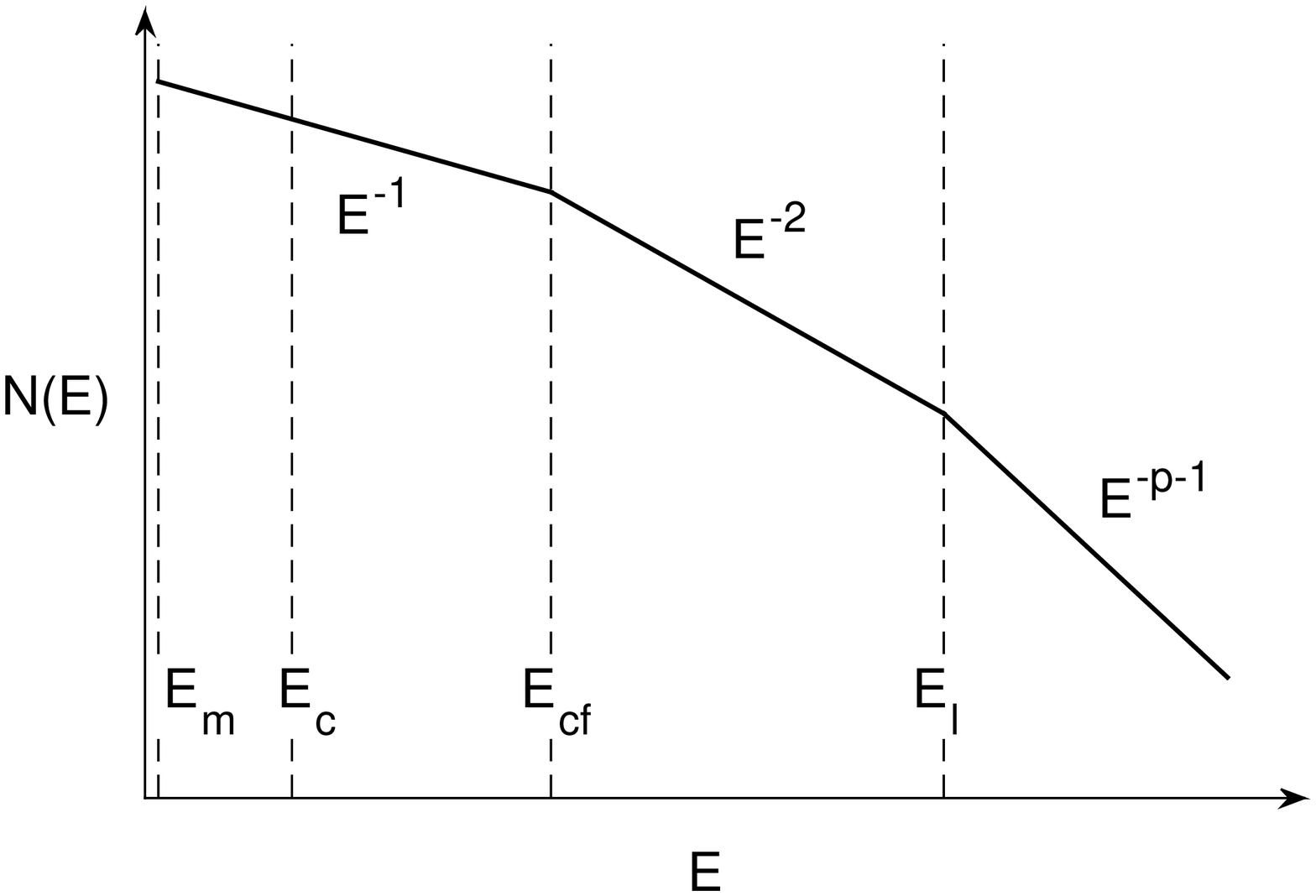}\label{fig: fac2}}
\subfigure[Case (\romannumeral3)]{
   \includegraphics[width=7.5cm]{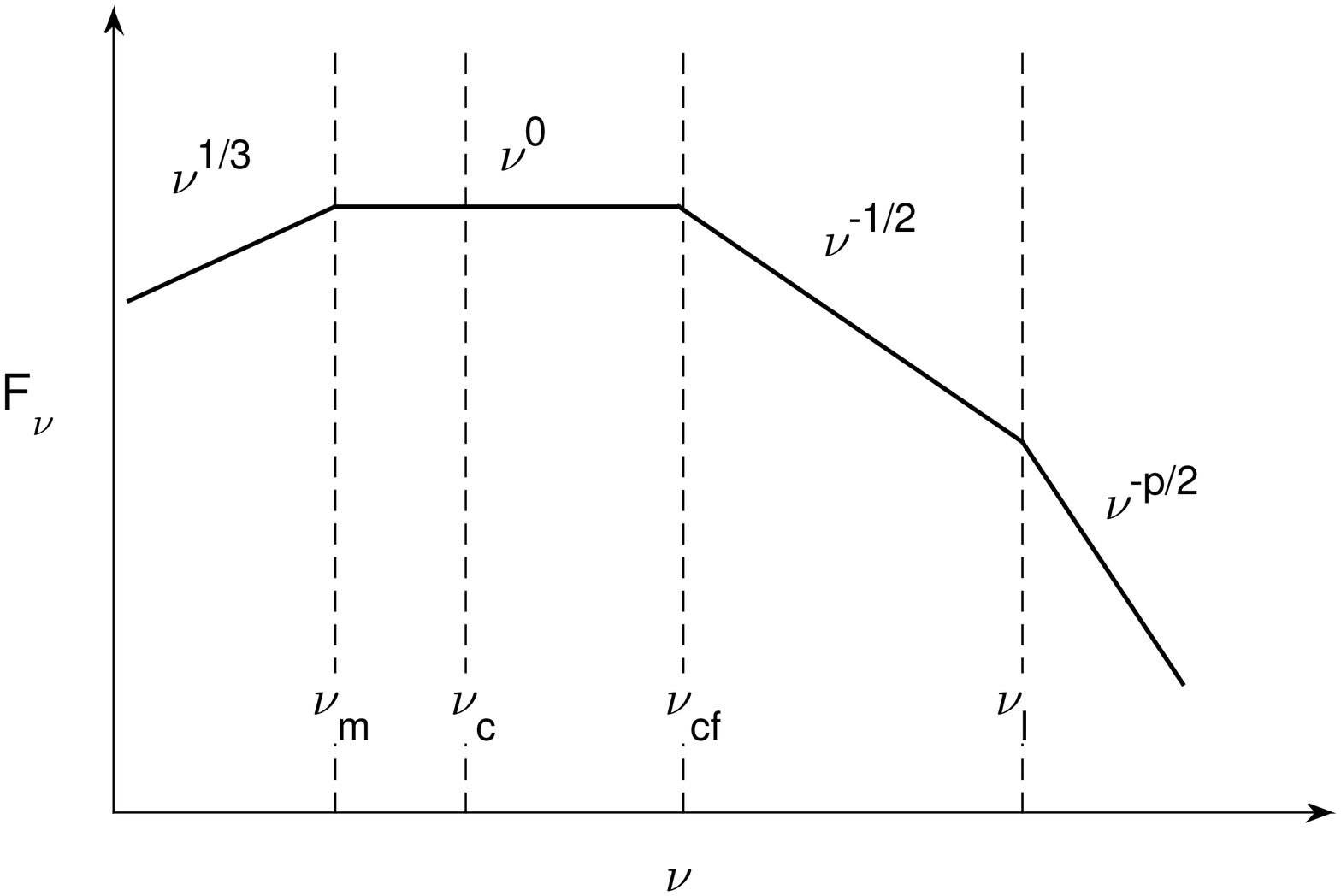}\label{fig: fac2s}}   
\subfigure[Case (\romannumeral4)]{
   \includegraphics[width=7.5cm]{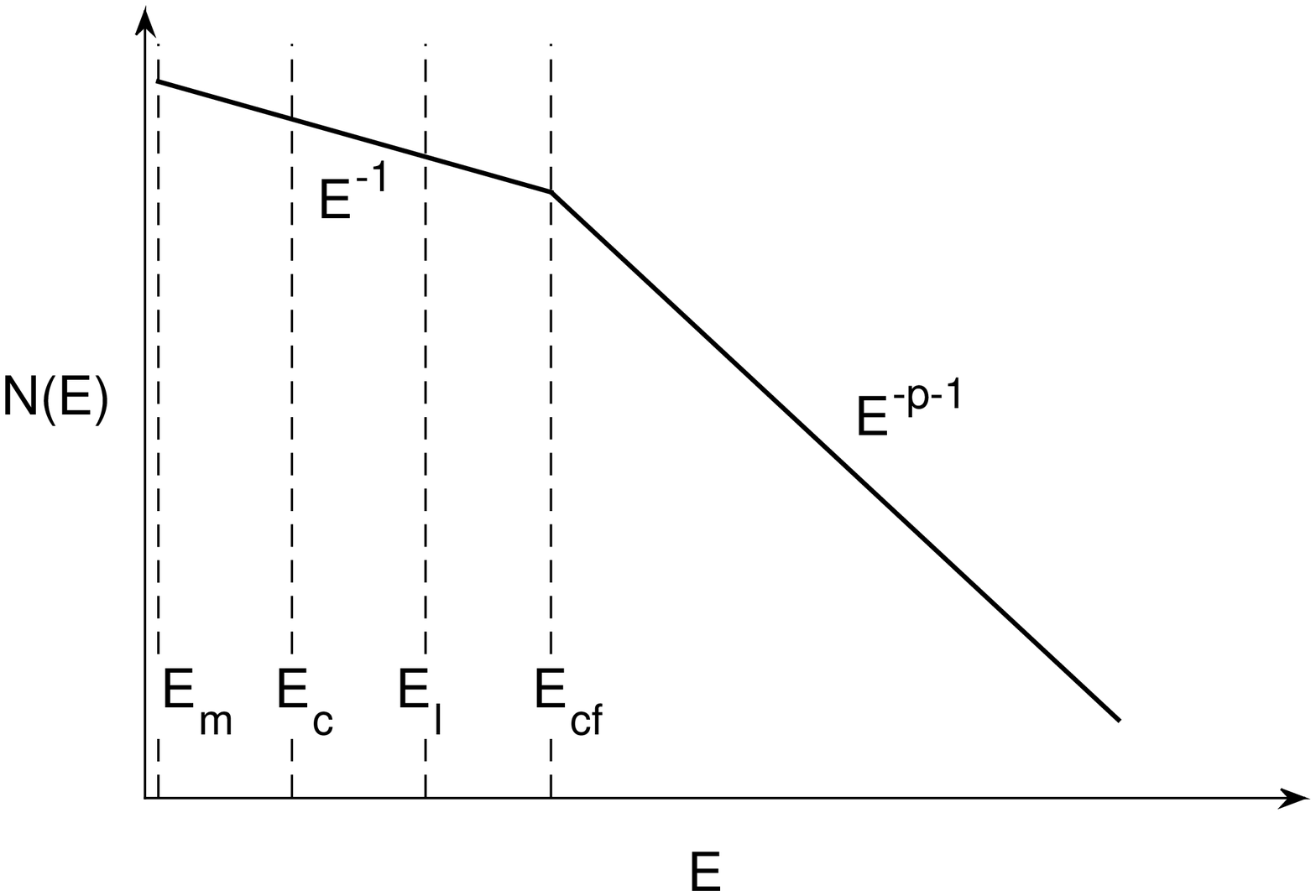}\label{fig: fac1}}
\subfigure[Case (\romannumeral4)]{
   \includegraphics[width=7.5cm]{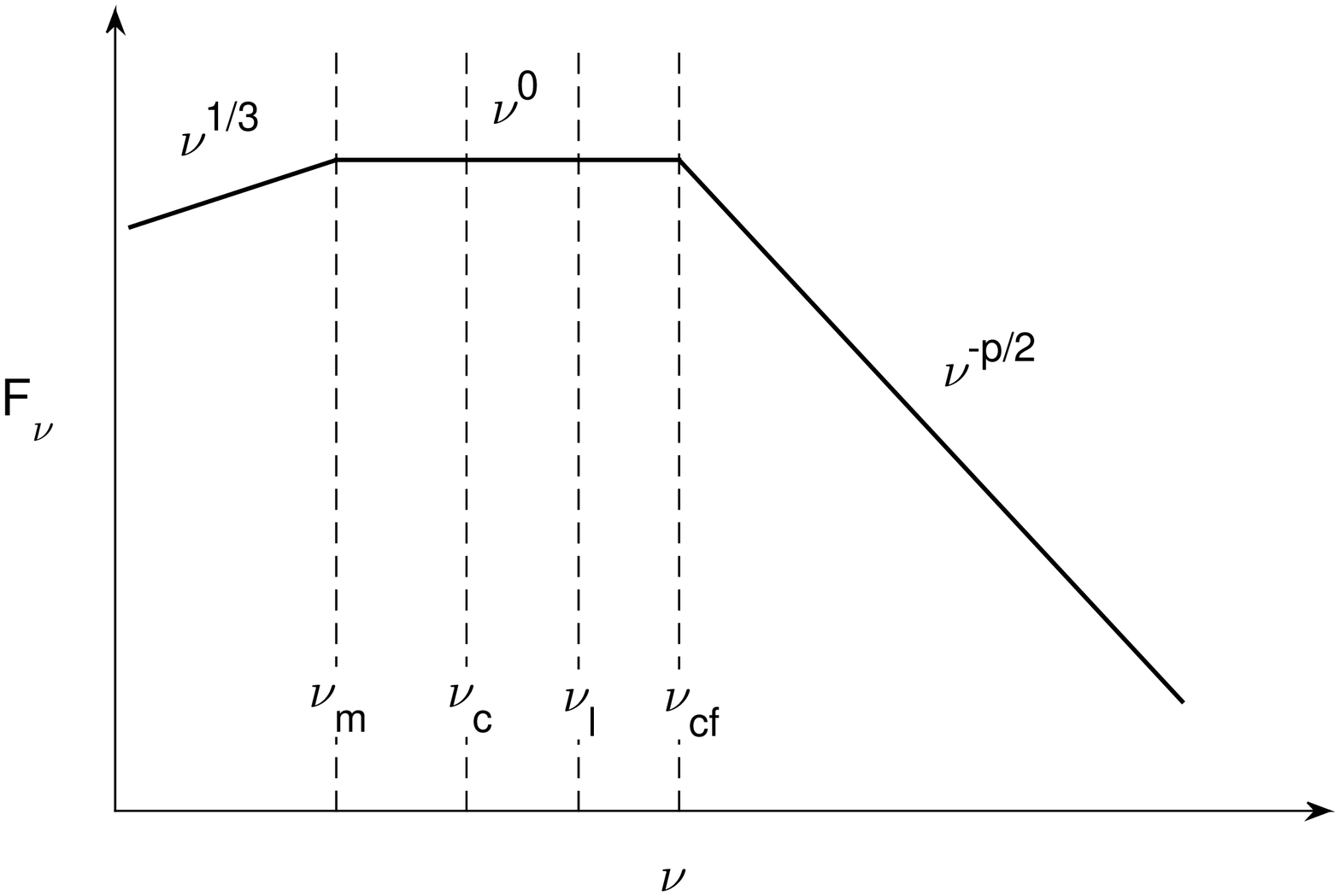}\label{fig: fac1s}}   

\caption{The electron energy spectrum ((a) and (c)) and the corresponding synchrotron spectrum ((b) and (d)) in Case (\romannumeral3)
and in Case (\romannumeral4). From \cite{XYZ18}.  }
\label{fig: fast2}
\end{figure}

In the energy range $E < E_\text{cf}$, 
the ASA dominates over the synchrotron cooling and leads to the electron energy spectrum as shown in Eq. \eqref{eq: fism}. 
At higher energies, the synchrotron cooling plays a dominant role in shaping the electron energy spectrum. 
Depending on the relation between $E_l$, $E_\text{cf}$, and $E_c$, 
where $E_c = 1/\beta t$ is the critical cooling energy  
\cite{Sar98}, 
the electron spectrum and the corresponding synchrotron spectrum 
exhibit different forms \cite{XYZ18}, 
as illustrated in Figs. \ref{fig: fast} and \ref{fig: fast2}. 
The asymptotic functional forms of the flux in different cases are as follows:

Case (\romannumeral1) $E_l < E_\text{cf} < E_c$,
\begin{subnumcases}
 {F_\nu=}
F_{\nu,\text{max}} \Big(\frac{\nu}{\nu_m}\Big)^\frac{1}{3}  ,~~~~~~~~~~~~~~~~~~~~~~~~\nu<\nu_m,\\
F_{\nu,\text{max}},~~~~~~~~~~~~~~~~~~~~~~~~~~~~\nu_m < \nu <\nu_\text{cf}, \\
F_{\nu,\text{max}} \Big(\frac{\nu}{\nu_\text{cf}}\Big)^{-\frac{p-1}{2}} ,~~~~~~~~~~\nu_\text{cf} < \nu < \nu_c, \\
F_{\nu,\text{max}} \Big(\frac{\nu_c}{\nu_\text{cf}}\Big)^{-\frac{p-1}{2}} \Big(\frac{\nu}{\nu_c}\Big)^{-\frac{p}{2}} ,~~~~~~\nu_c < \nu.
\end{subnumcases}
Case (\romannumeral2) $E_l < E_c < E_\text{cf}$,
\begin{subnumcases}
 {F_\nu=}
F_{\nu,\text{max}} \Big(\frac{\nu}{\nu_m}\Big)^\frac{1}{3}  ,~~~~~~~~~~~~~~~~~~~~~~~~\nu<\nu_m,\\
F_{\nu,\text{max}},~~~~~~~~~~~~~~~~~~~~~~~~~~~~\nu_m < \nu <\nu_\text{cf}, \\
F_{\nu,\text{max}} \Big(\frac{\nu}{\nu_\text{cf}}\Big)^{-\frac{p}{2}}, ~~~~~~~~~~~~~~~~~~~~~~\nu_\text{cf} < \nu.
\end{subnumcases}
Case (\romannumeral3) $E_c < E_\text{cf} < E_l$,
\begin{subnumcases}
 {F_\nu=}
F_{\nu,\text{max}} \Big(\frac{\nu}{\nu_m}\Big)^\frac{1}{3}  ,~~~~~~~~~~~~~~~~~~~~~~~~\nu<\nu_m,\\
F_{\nu,\text{max}},~~~~~~~~~~~~~~~~~~~~~~~~~~~~\nu_m < \nu <\nu_\text{cf}, \\
F_{\nu,\text{max}} \Big(\frac{\nu}{\nu_\text{cf}}\Big)^{-\frac{1}{2}} ,~~~~~~~~~~~~~~\nu_\text{cf} < \nu < \nu_l, \\
F_{\nu,\text{max}} \Big(\frac{\nu_l}{\nu_\text{cf}}\Big)^{-\frac{1}{2}} \Big(\frac{\nu}{\nu_l}\Big)^{-\frac{p}{2}} ,~~~~~~~~~~\nu_l < \nu.
\end{subnumcases}
Case (\romannumeral4) $E_c < E_l < E_\text{cf}$,
\begin{subnumcases}
 {F_\nu=\label{eq: fcc1}}
F_{\nu,\text{max}} \Big(\frac{\nu}{\nu_m}\Big)^\frac{1}{3}  ,~~~~~~~~~~~~~~~~~~~~~~~~\nu<\nu_m,\\
F_{\nu,\text{max}},~~~~~~~~~~~~~~~~~~~~~~~~~~~~\nu_m < \nu <\nu_\text{cf}, \\
F_{\nu,\text{max}} \Big(\frac{\nu}{\nu_\text{cf}}\Big)^{-\frac{p}{2}} ,~~~~~~~~~~~~~~~~~~~~~~~\nu_\text{cf} < \nu.
\end{subnumcases}

We note that the low-frequency tail $F_\nu \propto \nu^{1/3}$ comes from the synchrotron single-particle emission spectrum 
\cite{MR93,Kat94}.

\section{Application to the synchrotron spectra of GRB prompt emission}

The ASA can generally take place in MHD turbulence and dominate over other stochastic acceleration mechanisms, e.g., gyroresonance, 
when the pitch-angle scattering is inefficient. 
Therefore, we have explored the application of the ASA to the stochastic acceleration process 
in different astrophysical environments, which are both magnetized and turbulent. 
In this section and \S \ref{sec: pwn}, we will discuss the ASA in the context of GRBs and PWNe as examples.

The empirical Band spectrum 
\cite{Band93}
is usually used to describe the time-averaged synchrotron spectrum of GRB prompt emission. 
It is characterized by a low-energy spectral index $\alpha_s$, a break energy $E_b$, and a high-energy spectral index $\beta_s$. 
The distributions of $\alpha_s$ and $\beta_s$ are centered around $-1$ and $-2.25$, respectively, 
and $E_b$ is on the order of $100$ keV
\cite{Pre00}.
The above spectral features, especially the hard low-energy spectrum, cannot be well explained by either the thermal model
\cite{Bel10,Deng14} 
or the synchrotron model 
\cite{Ree92,Kat94,Tav96}
of GRBs.

In \cite{XZg17,XYZ18},
we investigated the ASA in the magnetized and turbulent GRB outflow 
\cite{Zh11,Deng15}
and found that it can naturally account for the hard low-energy spectrum of GRB prompt emission. 
The synchrotron spectrum in both Case (\romannumeral2) and Case (\romannumeral4) (\S \ref{sec: ssasa}) agrees with the Band spectrum.  
As a result of the ASA, there is 
$N(\nu) \propto F_\nu / \nu \propto \nu^{-1}$ in the frequency range $(\nu_m, \nu_\text{cf})$. 
The high-energy synchrotron spectrum is related to the injected electron distribution, and 
$\beta_s = -2.25$ corresponds to $p = 2.5$, which can be explained by other first-order Fermi acceleration processes, 
such as the shock acceleration and the reconnection acceleration 
\cite{DeG05}. 
$E_b$ is related to $E_\text{cf}$. The photon energy corresponding to $E_\text{cf}$ in the observer frame is 
\cite{XYZ18},
\begin{equation}
   E_{s,\text{cf,obs}} \simeq 385 \text{keV}~ \Big(\frac{1+z}{2}\Big)^{-1} \Gamma_2^{-0.2} L_{52}^{0.6} r_{15}^{-1.2},
\end{equation}
where $z$ is the redshift, $\Gamma = 100 \Gamma_2$ is the bulk Lorentz factor, 
$L = 10^{52}$ erg s$^{-1} L_{52}$ is the total isotropic luminosity, and 
$r = 10^{15}$ cm$r_{15}$ is the radius of the emission region
\cite{Zh11}. 
It has the same order of magnitude as $E_b$ that is indicated by observations.

\section{Application to the synchrotron spectra of PWNe}
\label{sec: pwn}

\begin{figure}[h]
\centering
\subfigure[]{
   \includegraphics[width=7cm]{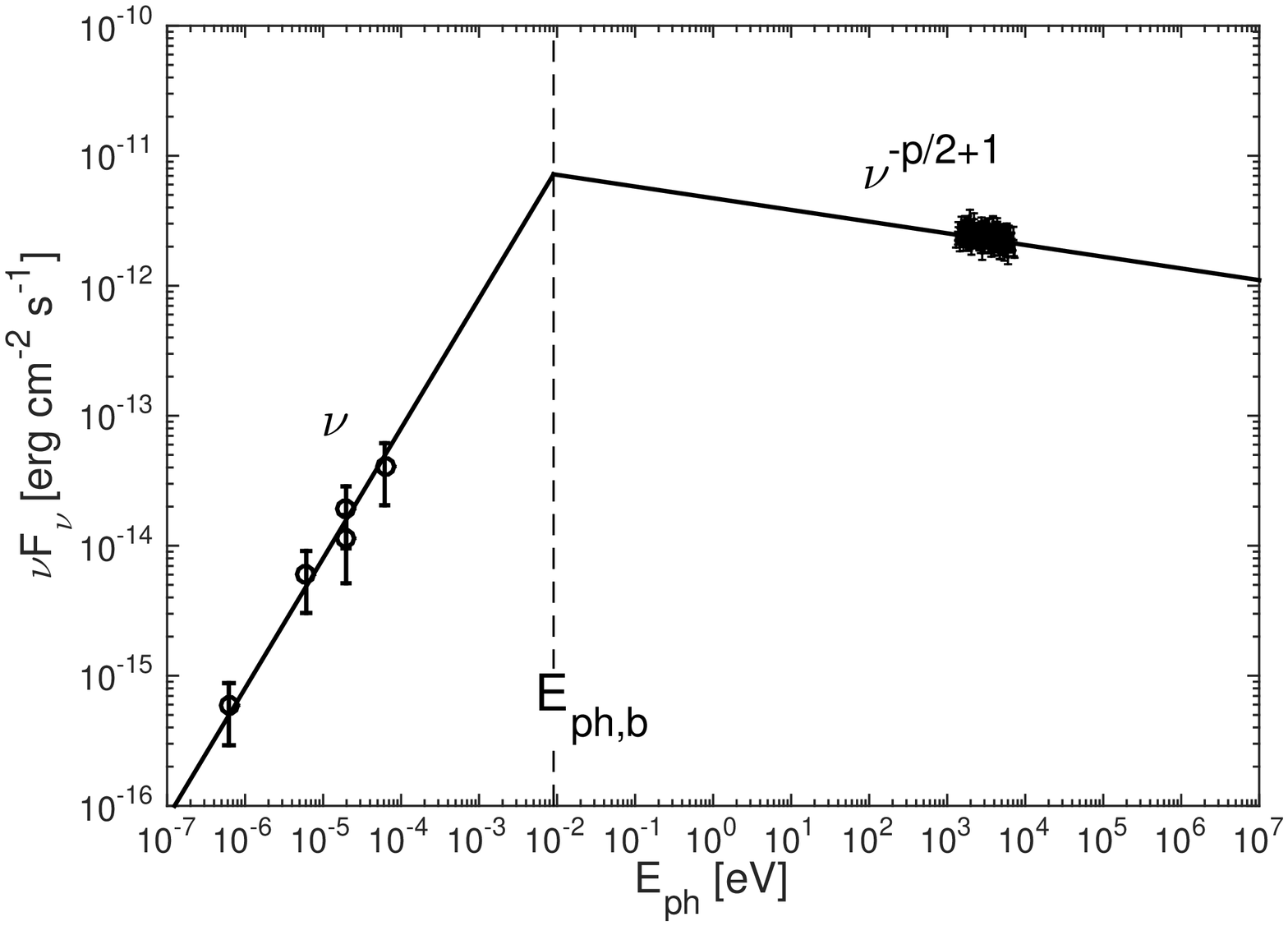}\label{fig:mou}}   
\subfigure[]{
   \includegraphics[width=7cm]{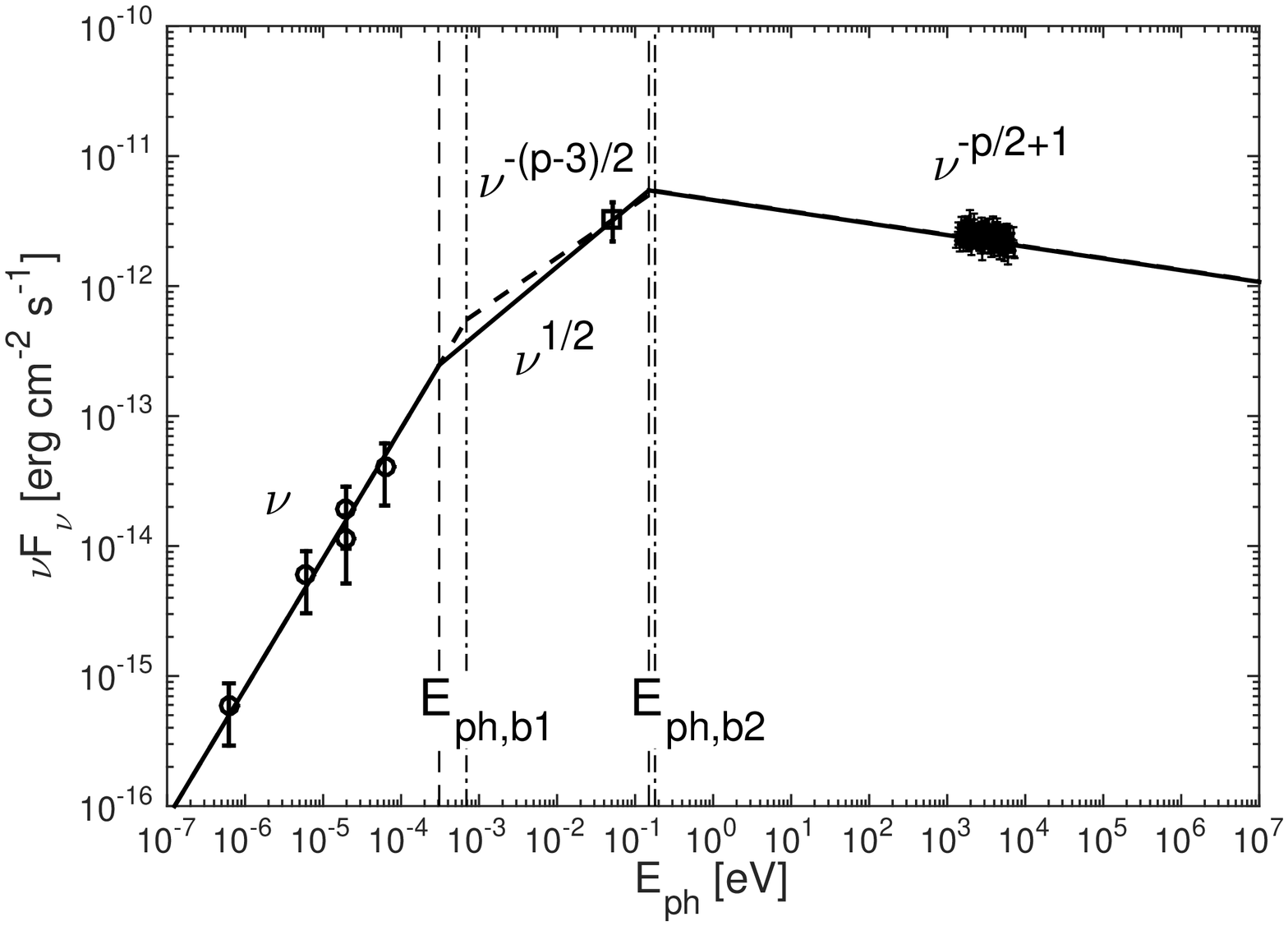}\label{fig:mou2}}   
\caption{Comparison between analytical spectral shapes and observational data for the Mouse PWN. 
Circles, square, and dots represent radio, infrared, and X-ray data, respectively.
See the original figure and references in 
\cite{Xu19}.}
\label{fig: mouse}
\end{figure}

\begin{figure}[h]
\centering
\subfigure[]{
   \includegraphics[width=7cm]{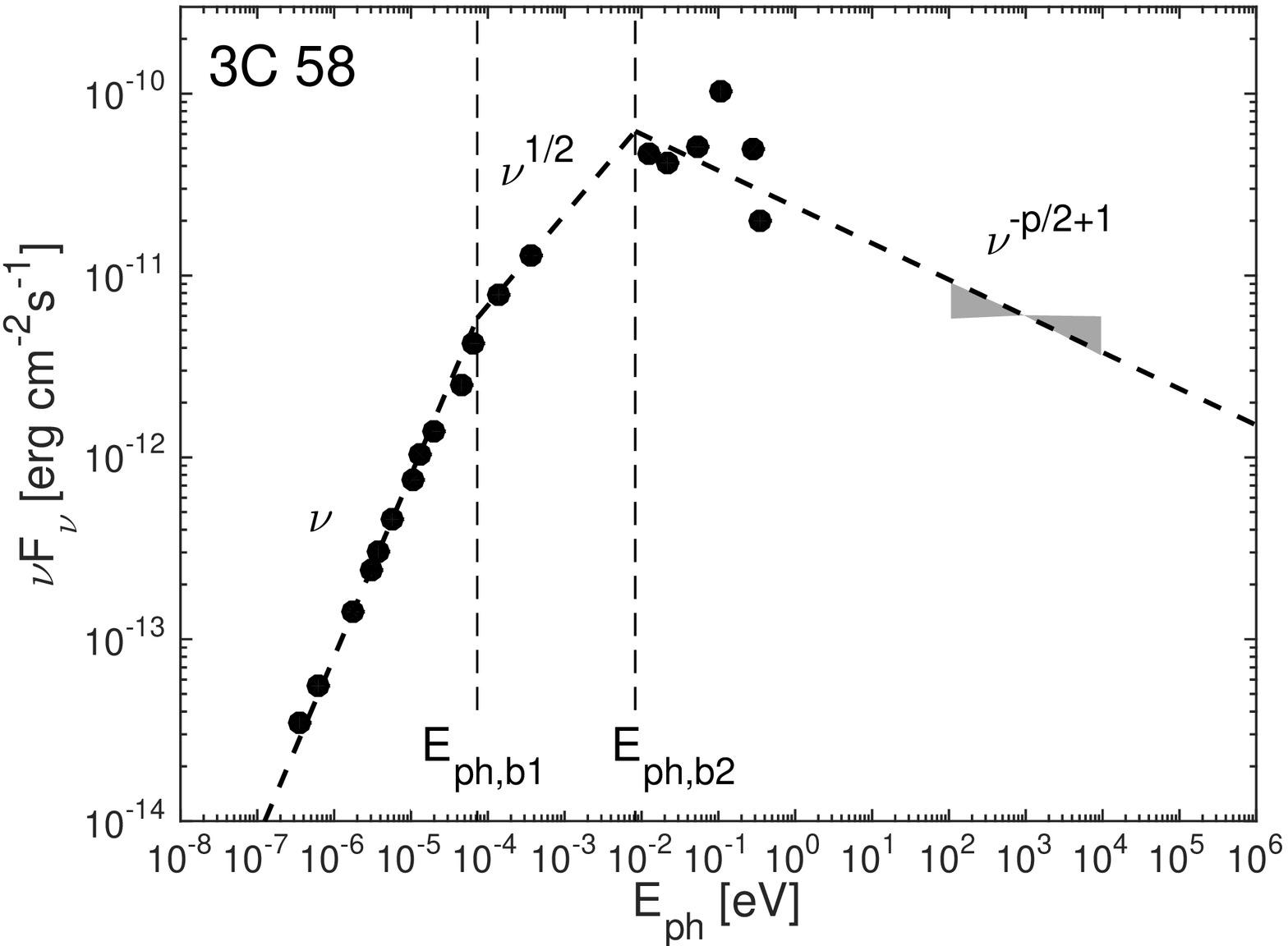}\label{fig: 3c}}  
\subfigure[]{
   \includegraphics[width=7cm]{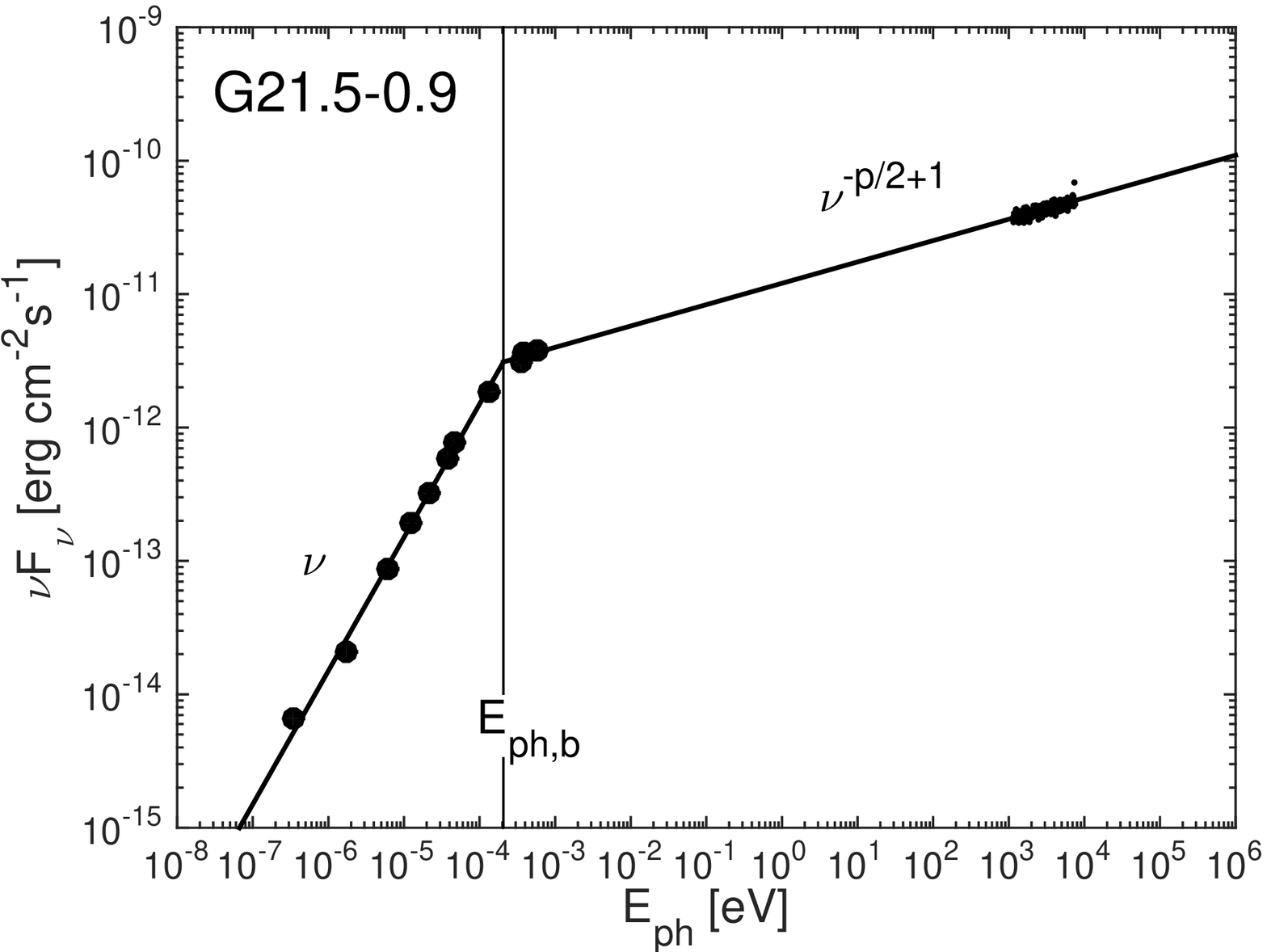}\label{fig: g21}}    
\subfigure[]{
   \includegraphics[width=7cm]{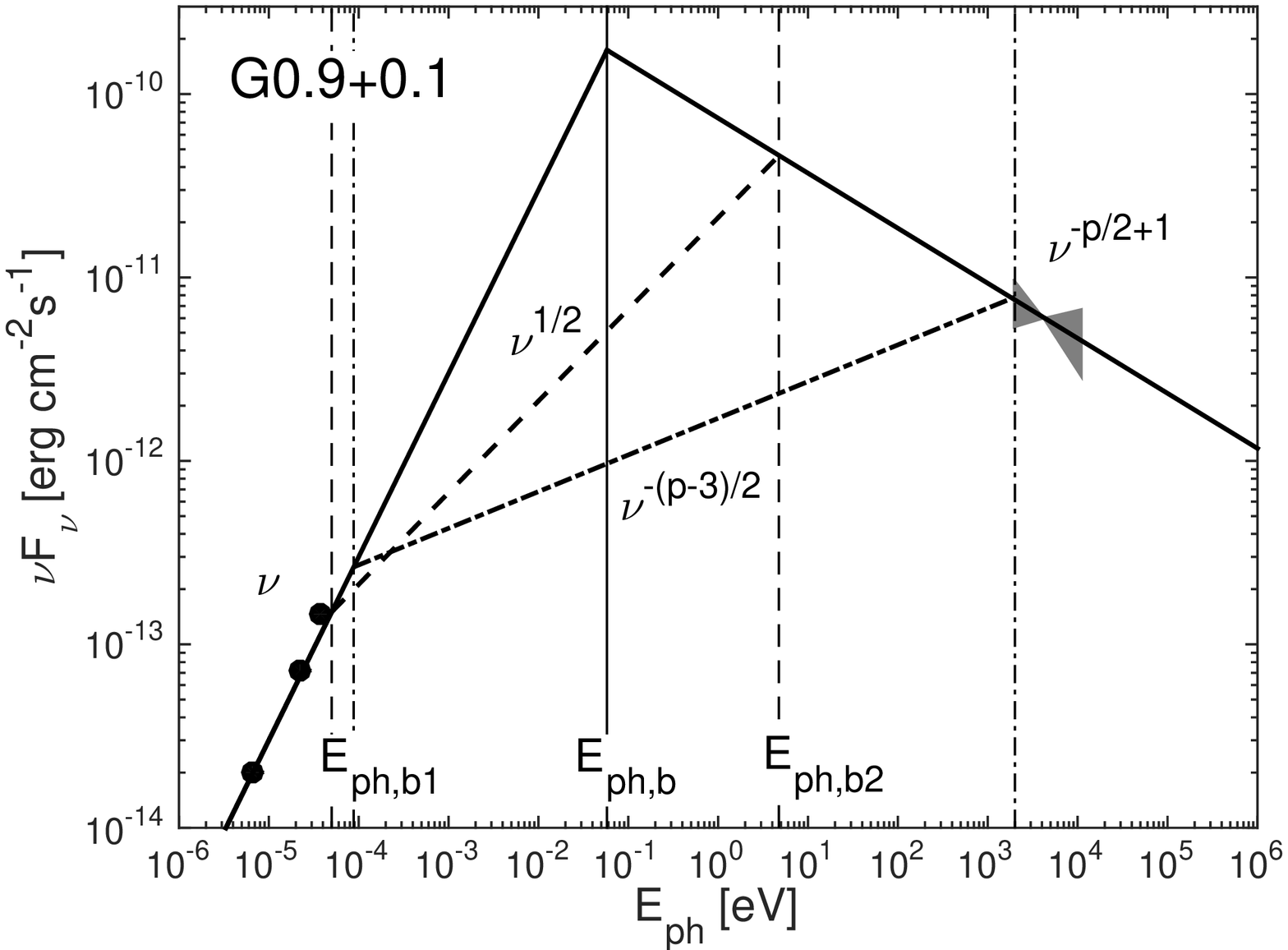}\label{fig: g09}}      
\subfigure[]{
   \includegraphics[width=7cm]{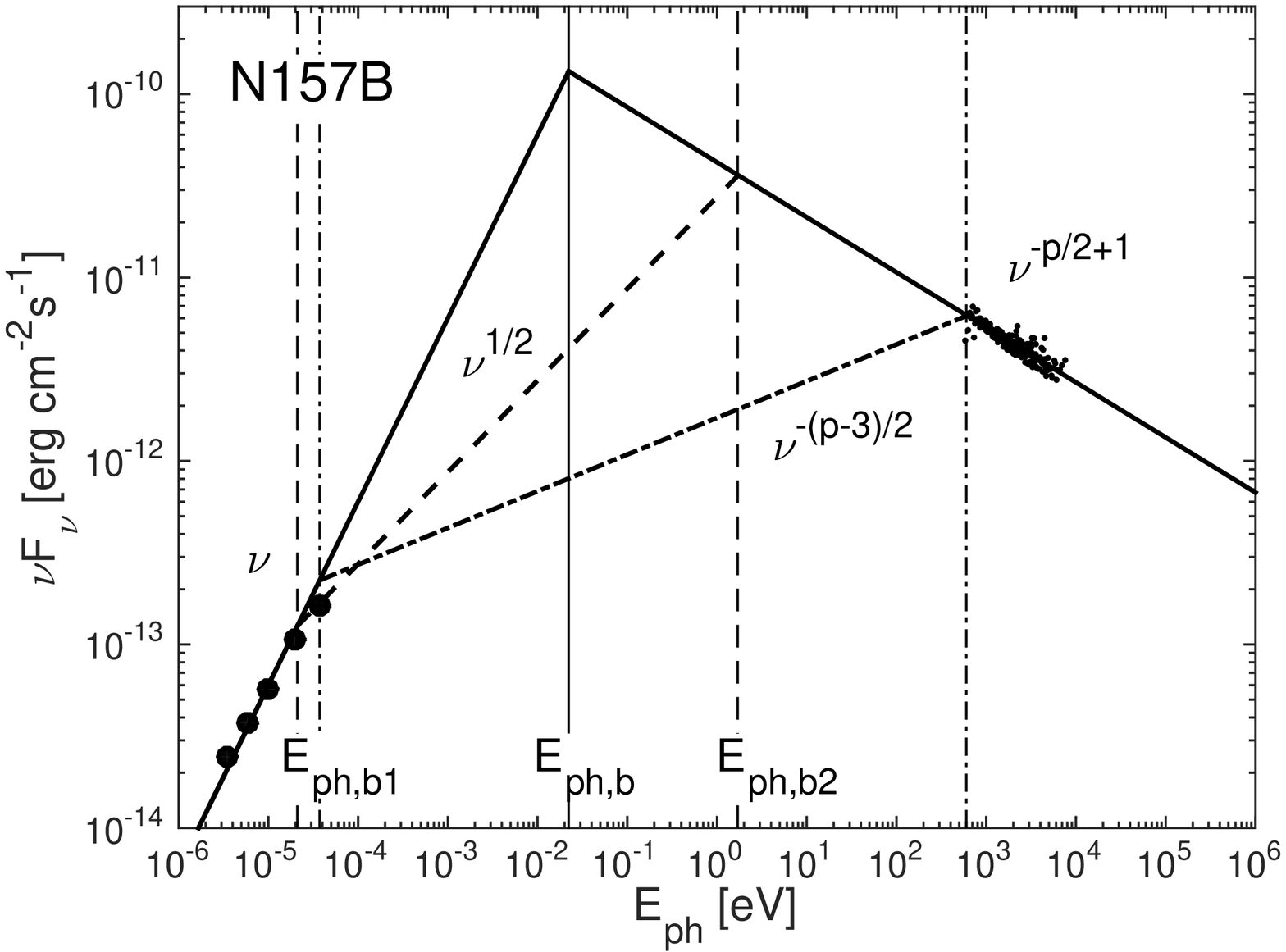}\label{fig: n157}}      
\caption{ Comparison between analytical spectral shapes and observational data for different PWNe. 
See the original figures and references in 
\cite{Xu19}.}
\label{fig: colle}
\end{figure}

\begin{figure}[h]
\includegraphics[width=9cm]{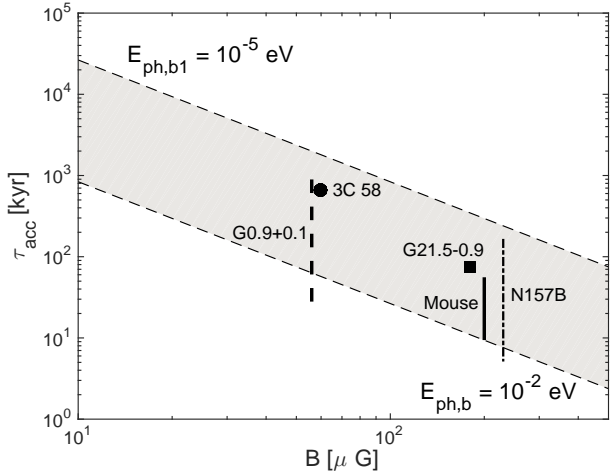}
\begin{minipage}[b]{14pc}
\caption{ $\tau_\text{acc}$ vs. $B_\text{eq}$ of the PWNe, 
where $B_\text{eq}$ is the magnetic field strength 
under the assumption of equipartition between particle and magnetic energies. 
The shaded region is bounded by Eq. \eqref{eq: parang}. 
From \cite{Xu19}.}
\label{fig: sum}
\end{minipage}
\end{figure}

In \cite{Xu19}, 
we applied the ASA to explaining the synchrotron spectra of magnetized and turbulent PWNe and found a good agreement between 
the analytical spectral shapes (\S \ref{sec: ssasa}) and the observed broad-band synchrotron spectra of different PWNe
(see Figs. \ref{fig: mouse} and \ref{fig: colle}). 
Moreover, the observed spectral break is related to $E_\text{cf}$ and can be used to constrain  the acceleration
timescale of the ASA $\tau_\text{acc} = 1/a_2$.
If the synchrotron spectrum of a PWN falls in Case (\romannumeral2) or Case (\romannumeral4), 
there is 
\begin{equation}\label{eq: cobss}
    \Big(\frac{\tau_\text{acc} }{1~\text{kyr}}\Big) \Big(\frac{B}{100 ~\mu \text{G}}\Big)^{\frac{3}{2}}
     = 26.5   \Big(\frac{\Gamma}{1}\Big)^{\frac{1}{2}}  \Big(\frac{E_\text{ph,b}}{ 10^{-2}~  \text{eV}}\Big)^{-\frac{1}{2}} , 
\end{equation}
where $B$ is the magnetic field strength in the PWN, 
$\Gamma$ is the bulk Lorentz factor of the mildly relativistic flow in the PWN, 
and $E_\text{ph,b}$ is the observed energy break of synchrotron spectrum. 
If the synchrotron spectrum of a PWN falls in Case (\romannumeral1) or Case (\romannumeral3), 
there is 
\begin{equation}\label{eq: coattw}
   \Big( \frac{\tau_\text{acc} }{1~\text{kyr}} \Big) \Big(\frac{B}{100 ~\mu \text{G}}\Big)^{\frac{3}{2}}
     = 837.0   \Big(\frac{\Gamma}{1}\Big)^{\frac{1}{2}}  \Big(\frac{E_\text{ph,b1}}{ 10^{-5}~  \text{eV}}\Big)^{-\frac{1}{2}}  ,
\end{equation}
where $E_\text{ph,b1}$ is the observed energy break at a lower energy (see Fig. \ref{fig:mou2}).
In the case when the spectral shape in the infrared band cannot be well determined, we have 
\begin{equation}\label{eq: parang}
26.5   \Big(\frac{\Gamma}{1}\Big)^{\frac{1}{2}}  \Big(\frac{E_\text{ph,b}}{ 10^{-2}~  \text{eV}}\Big)^{-\frac{1}{2}}  \leq
\Big( \frac{\tau_\text{acc} }{1~\text{kyr}} \Big) \Big(\frac{B}{100 ~\mu \text{G}}\Big)^{\frac{3}{2}} 
  \leq 837.0   \Big(\frac{\Gamma}{1}\Big)^{\frac{1}{2}}  \Big(\frac{E_\text{ph,b1}}{ 10^{-5}~  \text{eV}}\Big)^{-\frac{1}{2}}  ,
\end{equation}
as illustrated in Fig. \ref{fig: sum}.

\section{Summary}

As a new stochastic acceleration mechanism, the ASA arises from the basic dynamics of MHD turbulence involving both 
turbulent dynamo and turbulent reconnection of magnetic fields. 
Different from other stochastic acceleration mechanisms, it is highly efficient as particles undergo 
the first-order Fermi process within individual turbulent eddies. 
It is also not subject to the turbulence anisotropy effect, which makes the gyroresonance with Alfv\'{e}nic turbulence inefficient.

The ASA naturally hardens the injected electron spectrum and results in a hard electron spectrum in the energy range dominated by the ASA. 
Under the effects of both ASA and radiation cooling, 
the electron spectrum exhibits a broken power-law shape. 
The resulting synchrotron spectrum well explains the synchrotron spectrum of GRB prompt emission, as well as the 
broad-band synchrotron spectrum of a PWN.

The ASA is a general acceleration mechanism in MHD turbulence. 
Besides GRBs and PWNe, the application of the ASA to other high-energy astrophysical environments, e.g., 
radio galaxies,
blazars,
will be investigated in our future work.

\section*{Acknowledgement}

I acknowledge the support for Program number HST-HF2-51400.001-A provided by NASA through a grant from the Space Telescope Science Institute, which is operated by the Association of Universities for Research in Astronomy, Incorporated, under NASA contract NAS5-26555.
I am grateful to Bing Zhang,
Yuan-Pei Yang,
Noel Klingler, and Oleg Kargaltsev 
for their contributions to our studies mentioned here. 
This paper is based on an invited talk that I have given at the 18th Annual International Astrophysics Conference.

\bibliography{iopart-num}

\providecommand{\newblock}{}
\begin{thebibliography}{10}
\expandafter\ifx\csname url\endcsname\relax
  \def\url#1{{\tt #1}}\fi
\expandafter\ifx\csname urlprefix\endcsname\relax\def\urlprefix{URL }\fi
\providecommand{\eprint}[2][]{\url{#2}}

\bibitem{Zh11}
{Zhang} B and {Yan} H 2011 {\em ApJ\/} {\bf 726} 90 (\textit{Preprint}
  \eprint{1011.1197})

\bibitem{Deng15}
{Deng} W, {Li} H, {Zhang} B and {Li} S 2015 {\em ApJ\/} {\bf 805} 163
  (\textit{Preprint} \eprint{1501.07595})

\bibitem{Laz18}
{Lazarian} A, {Zhang} B and {Xu} S 2018 {\em arXiv: 1801.04061\/}
  (\textit{Preprint} \eprint{1801.04061})

\bibitem{XY13}
{Xu} S and {Yan} H 2013 {\em ApJ\/} {\bf 779} 140 (\textit{Preprint}
  \eprint{1307.1346})

\bibitem{Xuc16}
{Xu} S, {Yan} H and {Lazarian} A 2016 {\em ApJ\/} {\bf 826} 166
  (\textit{Preprint} \eprint{1506.05585})

\bibitem{Xult}
{Xu} S and {Lazarian} A 2018 {\em ApJ\/} {\bf 868} 36 (\textit{Preprint}
  \eprint{1810.07726})

\bibitem{GS95}
{Goldreich} P and {Sridhar} S 1995 {\em ApJ\/} {\bf 438} 763--775

\bibitem{LV99}
{Lazarian} A and {Vishniac} E~T 1999 {\em ApJ\/} {\bf 517} 700--718
  (\textit{Preprint} \eprint{arXiv:astro-ph/9811037})

\bibitem{CV00}
{Cho} J and {Vishniac} E~T 2000 {\em ApJ\/} {\bf 539} 273--282
  (\textit{Preprint} \eprint{arXiv:astro-ph/0003403})

\bibitem{MG01}
{Maron} J and {Goldreich} P 2001 {\em ApJ\/} {\bf 554} 1175--1196
  (\textit{Preprint} \eprint{arXiv:astro-ph/0012491})

\bibitem{CLV_incomp}
{Cho} J, {Lazarian} A and {Vishniac} E~T 2002 {\em ApJ\/} {\bf 564} 291--301
  (\textit{Preprint} \eprint{arXiv:astro-ph/0105235})

\bibitem{XL16}
{Xu} S and {Lazarian} A 2016 {\em ApJ\/} {\bf 833} 215 (\textit{Preprint}
  \eprint{1608.05161})

\bibitem{BruL16}
{Brunetti} G and {Lazarian} A 2016 {\em MNRAS\/} {\bf 458} 2584--2595
  (\textit{Preprint} \eprint{1603.00458})

\bibitem{YL02}
{Yan} H and {Lazarian} A 2002 {\em Physical Review Letters\/} {\bf 89} B1102+
  (\textit{Preprint} \eprint{arXiv:astro-ph/0205285})

\bibitem{Band93}
{Band} D, {Matteson} J, {Ford} L, {Schaefer} B, {Palmer} D, {Teegarden} B,
  {Cline} T, {Briggs} M, {Paciesas} W, {Pendleton} G, {Fishman} G,
  {Kouveliotou} C, {Meegan} C, {Wilson} R and {Lestrade} P 1993 {\em ApJ\/}
  {\bf 413} 281--292

\bibitem{XZg17}
{Xu} S and {Zhang} B 2017 {\em ApJL\/} {\bf 846} L28 (\textit{Preprint}
  \eprint{1708.08029})

\bibitem{XYZ18}
{Xu} S, {Yang} Y~P and {Zhang} B 2018 {\em ApJ\/} {\bf 853} 43
  (\textit{Preprint} \eprint{1711.03943})

\bibitem{Xu19}
{Xu} S, {Klingler} N, {Kargaltsev} O and {Zhang} B 2019 {\em ApJ\/} {\bf 872}
  10 (\textit{Preprint} \eprint{1812.10827})

\bibitem{Sar98}
{Sari} R, {Piran} T and {Narayan} R 1998 {\em ApJL\/} {\bf 497} L17--L20
  (\textit{Preprint} \eprint{astro-ph/9712005})

\bibitem{MR93}
{Meszaros} P and {Rees} M~J 1993 {\em ApJL\/} {\bf 418} L59 (\textit{Preprint}
  \eprint{astro-ph/9309011})

\bibitem{Kat94}
{Katz} J~I 1994 {\em ApJL\/} {\bf 432} L107--L109 (\textit{Preprint}
  \eprint{astro-ph/9312034})

\bibitem{Pre00}
{Preece} R~D, {Briggs} M~S, {Mallozzi} R~S, {Pendleton} G~N, {Paciesas} W~S and
  {Band} D~L 2000 {\em ApJS\/} {\bf 126} 19--36 (\textit{Preprint}
  \eprint{astro-ph/9908119})

\bibitem{Bel10}
{Beloborodov} A~M 2010 {\em MNRAS\/} {\bf 407} 1033--1047 (\textit{Preprint}
  \eprint{0907.0732})

\bibitem{Deng14}
{Deng} W and {Zhang} B 2014 {\em ApJ\/} {\bf 785} 112 (\textit{Preprint}
  \eprint{1402.5364})

\bibitem{Ree92}
{Rees} M~J and {Meszaros} P 1992 {\em MNRAS\/} {\bf 258} 41P--43P

\bibitem{Tav96}
{Tavani} M 1996 {\em ApJ\/} {\bf 466} 768

\bibitem{DeG05}
{de Gouveia dal Pino} E~M and {Lazarian} A 2005 {\em A\&A\/} {\bf 441} 845--853

\end{thebibliography}
\end{document}